\documentclass[prb,twocolumn,showpacs,preprintnumbers,amsmath,amssymb]{revtex4}

\usepackage[dvips]{graphicx}
\usepackage{graphicx}% Include figure files
\usepackage{dcolumn}% Align table columns on decimal point
\usepackage{bm}% bold math
\usepackage{amsmath,amssymb}
\usepackage{color}
\usepackage{ulem}

\begin{document}

\textbf{}
\title{Temperature range of superconducting fluctuations above $T_c$ 
in $\rm YBa_2Cu_3O_{7-\delta}$ single crystals}

\author{M.~S.~Grbi\'c$^1$}
\author{M.~Po\v zek$^1$} 
\author{D.~Paar$^1$}
\author{V.~Hinkov$^2$}
\author{M.~Raichle$^2$}
\author{D.~Haug$^2$}
\author{B.~Keimer$^2$}
\author{N.~Bari\v si\'c$^{3}$}
\author{A.~Dul\v ci\'c$^1$}

\email{adulcic@phy.hr}
\affiliation{$^1$Department of Physics, Faculty of Science, University of Zagreb, P. O. Box 331, HR-10002 Zagreb, Croatia\\
$^2$Max-Planck-Institut f\"ur Festk\"orperforschung, D-70569 Stuttgart, Germany\\
$^3$1.~Physikalisches Institut, Universit\"at Stuttgart, D-70550 Stuttgart, Germany\\}

\begin{abstract}
Microwave absorption measurements in magnetic fields from zero up to 16 T were used to determine the temperature range of superconducting fluctuations above the superconducting critical temperature $T_c$ in  $\rm YBa_2Cu_3O_{7-\delta}$. Measurements were performed on deeply underdoped, slightly underdoped, and overdoped single crystals. The temperature range  of the superconducting fluctuations above $T_c$ is determined by an  experimental method which is free from arbitrary assumptions about subtracting the nonsuperconducting contributions to the total measured signal, and/or theoretical models to extract the unknown parameters. The superconducting fluctuations are detected in the $ab-$plane, and $c-$axis conductivity, by identifying the onset temperature $T'$. Within the sensitivity of the method, this fluctuation regime is found only within a fairly narrow region above $T_c$. Its width increases from 7 K in the overdoped sample ($T_c = 89~\rm K$), to at most 23 K in the deeply underdoped sample ($T_c = 57~\rm K$), so that $T'$ falls well below the pseudogap temperature $T^{*}$. Implications of these findings are discussed in the context of other experimental probes of superconducting fluctuations in the cuprates.
\end{abstract}


\maketitle

\section{Introduction}

Superconducting fluctuations have attracted a great deal of attention in the research of high temperature superconductors (HTSC). The first reason was the challenge of observing the critical regime of fluctuations, which had been previously inaccessible in experiments with  classical low temperature superconductors. Due to the short coherence lengths, and high thermal energy $k_B T_c$, it was estimated from the Ginzburg criterion that critical fluctuations in HTSC could extend within 1~K, or more, around $T_c$.\cite{Lobb} Beyond the critical region, one expected a transition to Gaussian fluctuations, which are the lowest order fluctuation correction to the mean field theory.\cite{Skocpol, Tinkham} Indeed, critical fluctuations have been observed in HTSC by a number of experimental techniques, including penetration depth measurements \cite{Kamal}, thermal expansivity \cite{Pasler, Meingast}, two-coil inductive measurements \cite{Osborn}, and microwave measurements.\cite{Anlage, Booth, Nakielski, Waldram, Peligrad:04} The experimental evidence was consistent with the 3D XY universality class. The 3D character implies that the fluctuation correlations extend over many atomic layers along the $c$ axis. Even in strongly underdoped $\rm YBa_2Cu_3O_{7-\delta}$ (YBCO), which is extremely anisotropic, no evidence was found for the crossover to uncorrelated fluctuations in adjacent layers, i. e. the Kosterlitz-Thouless-Berezinsky (KTB) vortex-unbinding transition\cite{Broun} until the thickness of the sample became comparable to the fluctuation correlation length. \cite{Hetel} 

The second reason for the study of superconducting fluctuations in HTSC was related to the nature of the pseudogap, which is known to open in underdoped compounds at a temperature $T^*$, much above the superconducting transition temperature $T_c$.\cite{Timusk} The intriguing question was whether a highly sensitive experimental technique could detect the superconducting fluctuations extending as high above $T_c$ to almost reach the pseudogap temperature $T^*$. Observation of the superconducting fluctuations up to $T^*$ would indicate that the pseudogap is related to superconductivity. In such a scenario, the long-range coherence is lost at $T_c$ due to fluctuations of the phase of the superconducting order parameter.\cite{Emery, Emery2, Lee, Franz, Herbut, Anderson} The pseudogap region would then be marked by preformed Cooper pairs, and take the role of the precursor to the superconducting state. If, in contrast, one finds convincing experimental evidence that the onset temperature of the superconducting fluctuations falls well below $T^*$, in particular for underdoped HTSC, the nature of the pseudogap would appear to be unrelated to superconductivity. Instead, it could be explained by fluctuations of other order parameters, as proposed on theoretical grounds.\cite{Varma, Chakravarty, Sunko} 

Experiments probing magnetic correlations have recently uncovered evidence of electronic liquid-crystal states, \cite{Hinkov:07, Hinkov:08, Haug} and/or orbital-current order \cite{Fauque, Mook, Li, Sonier} in the pseudogap regime of underdoped HTSC. The onset of these ordering phenomena was found to coincide with $T^*$, as determined by the departure from linearity of the dc resistivity curve. However, the evidence presented in favor of new forms of magnetic order in the pseudogap region need not exclude the existence of some form of the superconducting order, or possible competition of the two.\cite{Kondo}. The extent of the superconducting fluctuations has to be firmly established using a sensitive and reliable experimental technique.

A major problem in most experimental techniques is to subtract properly a background signal unrelated to  superconductivity. For example, when analyzing the Nernst effect, one has to subtract the normal quasiparticle contribution,\cite{Xu, Wang:06} which may not obey the usual assumptions of the so-called Sondheimer cancellation.\cite{Behnia} Furthermore, it has also been shown that spin/charge density modulations, known as stripe order in underdoped HTSC, can cause a considerable contribution to the observed Nernst signal.\cite{Olivier} The conclusions on the extent of the superconducting fluctuations drawn from the Nernst effect measurements are still controversial,\cite{Li:10, Daou} and call for a complementary analysis by other experimental techniques.  

Early work on dc fluctuation conductivity in HTSC relied on the assumption that the normal state resistivity was linear, not only at temperatures far above $T_c$, but also closer to $T_c$ where the superconducting fluctuations appear.\cite{Ausloos, Hopfengartner, Costa, Han} The linearly extrapolated resistivity was used to calculate the normal conductivity $\sigma_n$, which was then subtracted from the total experimental conductivity to obtain the fluctuation conductivity alone. However, it was later shown that the linear normal state resistivity was not generic to HTSC, particularly not in underdoped compounds,\cite{Timusk, Takenaka} so that deviations from linearity could not be unambiguously ascribed to superconductivity.

In microwave measurements one can determine the complex conductivity $\tilde{\sigma} = \sigma_1 - i \sigma_2$,\cite{Anlage, Waldram, Peligrad:04, Corson} where the real part $\sigma_1$ includes both the normal conductivity $\sigma_n$ and the contribution from the superconducting fluctuations $\sigma_{1}^{\:'}$. Separating those two parts poses the same problem as in the dc conductivity analysis. The imaginary part $\sigma_2$ requires no subtractions, since it is entirely due to the appearance of superconductivity. However, it decays above $T_c$ much faster than the real part $\sigma_{1}^{\:'}$ of the fluctuation conductivity,\cite{Peligrad:03} and may become untraceable due to the signal noise. In order to deal with the problem, we have recently introduced a novel approach to microwave measurements in combination with an external magnetic field.\cite{Grbic} The microwave signal measured in 8 T field was subtracted from the one in zero field, and the difference was reliably attributed to the contribution of the superconducting fluctuations above $T_c$. In a nearly optimally doped $\rm HgBa_2CuO_{4+\delta}$ (Hg1201) high quality single crystal\cite{Barisic} with $T_c = 94.3~\rm K$, we could observe the superconducting fluctuations up to $T' = 105~\rm K$, which is only 10 K above $T_c$. This is far below the pseudogap temperature $T^* = 185~\rm K$ determined in this compound. 

In this paper, we take advantage of the microwave technique combined with the magnetic field to study the superconducting fluctuations in YBCO single crystals over a wide doping range. We have studied both overdoped and  underdoped samples (Table \ref{tb1}). For a short notation, we name these samples OD89, UD87, and UD57. The experimental arrangement allows for the determination of the superconducting  fluctuation conductivity separately in the $ab$ plane, and along the $c$ axis. Our results show that the region of superconducting fluctuations is fairly narrow. In the overdoped sample, it extends to only ~7 K above $T_c$, while in our most underdoped sample at most to ~23 K, which falls far below the pseudogap at that doping. Implications of these findings will be discussed.

\section{Experimental}

High-quality YBCO single crystals were grown by the solution growth method in the crystal growth group at the
Max-Planck-Institut f\"ur Festk\"orperforschung Stuttgart. In order to prepare samples with well-defined oxygen content the as-grown crystals were annealed in O$_2$ (OD89 and UD87) and synthetic air (UD57) respectively. The overdoped crystal was detwinned by application of uniaxial mechanical stress ($\sim 5 \times 10^{7}$ N/m$^2$) along the crystallographic $\langle 100 \rangle$ direction at elevated temperatures. Further details about the sample preparation are given in Ref. \onlinecite{Hinkov:04}.

%\begin{center}
\begin{table}[h]
\begin{ruledtabular}
\begin{tabular}{ccccccc}
Sample & Doping $p$ & $T_c$(K) & $\Delta T_c$(K) & Size (mm$^3$) & $N_{ab}$ & $N_{c}$\\
\hline
OD89 & 0.19 & 89.4 & 1.6  & 2.5$\times$1.9$\times$0.6 & 0.13 & 0.68\\
UD87 & 0.15 & 87.2 & 2  & 2.5$\times$1.95$\times$1  & 0.26 & 0.56\\
UD57 & 0.12 & 57.2 & 5  & 2.5$\times$1.6$\times$0.9 & 0.29 & 0.55\\
\end{tabular}
\end{ruledtabular}
\caption{Some physical properties of the measured samples. $N_{ab}$ and $N_{c}$ are the demagnetization factors\cite{Osborn45} for $H_{\rm mw}$ parallel to the $ab$ plane, and the $c$ axis, respectively.} 
\label{tb1}
\end{table}
%\end{center}

Table \ref{tb1} summarizes the main characteristics of the measured samples. The hole doping level per planar Cu ion $p$ was extracted from the known doping dependence of the room-temperature thermoelectric power \cite{Tallon} and the lattice parameter $c$ (Ref. \onlinecite{Liang}). The width $\Delta T_c$ of the superconducting transition was estimated by the common 10 - 90 percent criterion of the transition curve measured by the microwave absorption described later in this paper. We may note that the same criterion applied in dc resistivity curves yields a smaller transition width. This comes out naturally because of the nonzero losses in microwave measurements even below $T_c$. Thus, the present estimate of $\Delta T_c$ is a rather conservative one. The precise value quoted for $T_c$ is not the simple midpoint in the transition curve measured by the microwave absorption, but was infered from the maximum of the real part $\sigma_1$ of the complex microwave conductivity as shown later in Section V. This value of $T_c$ was found to lie within the $\Delta T_c$ range, somewhat closer to the lower (10 percent) boundary. 

For the microwave measurements, the sample was mounted on one end of a sapphire holder, which serves as a cold finger. The assembly was then inserted into the microwave cavity so that the sample takes the center position inside. Outside of the microwave cavity, a heater-sensor block was mounted on the free end of the sapphire holder. The temperature could be controlled to 10 mK stability. The microwave cavity was made of copper so as to allow measurements in any external magnetic field with only a small change in 
its $Q$-factor. This small field dependent variation could be detected by measuring the empty cavity, and subtracted from the signal measured when the cavity was loaded with the sample. 

The geometry of the microwave cavity was carefully chosen as elliptic, so that the degeneracy of some orthogonal modes could be lifted. We used $_{e}\rm TE_{112}$, and $_{e}\rm TE_{211}$ modes operating at 13.14 GHz, and 15.15 GHz, respectively. Each of these modes has an  antinode of the microwave magnetic field $H_{\rm mw}$ in the cavity center, where the sample is positioned. However, the polarization of $H_{\rm mw}$ is along the cavity axis in the $_{e}\rm TE_{211}$ mode, and perpendicular to this axis in the $_{e}\rm TE_{112}$ mode. Thus, with the same mounting and positioning of the sample in the cavity, we could carry out two measurements. We chose to mount the YBCO single crystal so that its $c$ axis was along the cavity axis (vertical in our experimental setup). Consequently, the induced microwave currents in the measurement with the $_{e}\rm TE_{211}$ mode were flowing as closed loops in the $ab$ plane, while the $_{e}\rm TE_{112}$ yielded microwave currents closing the loops by flowing partly in the $ab$ plane, and partly along the $c$ axis. The polarizations of $H_{\rm mw}$, and the induced current loops in the two respective modes are shown in the insets of Figs. \ref{SlikaOP} and \ref{SlikaUD}.

The microwave source was a Rohde Schwarz synthesizer whose frequency could be tuned to achieve resonance. An AFC circuit allowed tracking of the resonance as the temperature of the sample was swept, and the resonant frequency was measured by a microwave counter. On top of that, the microwaves were frequency-modulated at 991 Hz, with a frequency deviation which corresponds to the halfwidth of the Lorentzian representing the resonance curve of the microwave cavity loaded with the sample. Even harmonics of the output signal from the microwave cavity were used to determine very accurately the $Q$-factor of the cavity. The details of this method were published  elsewhere.\cite{Nebendahl} 

Data acquisition was made during a slow temperature sweep of the sample so that the microwave signal generator could alternate between the resonance frequencies of the two modes. Thus, we obtained the experimental curves in both modes with a single temperature sweep. This procedure ensures that the resulting data can be directly compared.

\section{Microwave absorption and superconducting fluctuations}

The results of microwave measurements are expressed through a change in the inverse value of the quality factor of the cavity loaded with the sample. In our experimental setup, the microwave cavity is kept always at a constant low temperature (e. g. 2~K) so that the power dissipation in its walls remains unchanged. The contribution of the unloaded cavity, $1/2Q_0$, is subtracted from the total experimental value $1/2Q_{exp}$ to obtain the contribution due to the sample itself $1/2Q = 1/2Q_{exp} - 1/2Q_0$. As the temperature of the sample is varied, one observes changes of $1/2Q$ as the absorption of microwaves is changed within the skin depth of the sample surface.

\subsection{Slightly underdoped sample}

In order to extract the contribution of the superconducting fluctuations from the overall signal, we have carried out measurements in zero magnetic field and in (dc) magnetic fields up to 16 T. The behavior of the superconducting fluctuations in applied magnetic fields has been previously studied by dc conductivity measurements,\cite{Axnas:98, Axnas:96, Holm:93} and compared to theoretical predictions.\cite{Dorin:93} Also, microwave conductivity measurements were used for studying the scaling properties in field and temperature.\cite{Ukrainczyk:95} In contrast to those studies, here we do not aim to study the  superconducting fluctuations in various applied fields, but apply a high enough field in order to suppress superconductivity at all temperatures above the zero field $T_c$. This serves us to determine the normal state background line without any extrapolation based on a theoretical model of the normal state behavior. In Section V, we will show more explicitly that 8 T field is already sufficient for this task, and 16 T field can further suppress the superconductivity a few degrees below the zero field $T_c$. 

\begin{figure}[t]
\includegraphics[width=8cm]{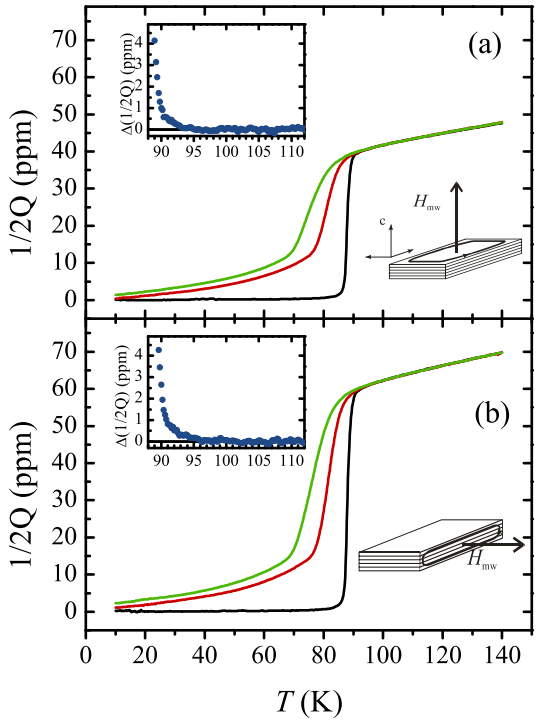}
\caption{(Color online) Microwave absorption in slightly underdoped YBCO ($T_c=87~\rm \,K$). External dc magnetic field is applied parallel to the $c$ axis (black: $B_{dc}=0$, red: $B_{dc}=8 \rm \, T$, green: $B_{dc}=16 \rm \, T$ ). (a) $H_{\rm mw}$ is in the $c$-direction so that the induced microwave currents flow in the $ab$ plane; (b) $H_{\rm mw}$ lies in the $ab$ plane so that the induced microwave currents flow partly in the $c$-direction and partly in the $ab$ plane. Insets show the difference in absorption between $B_{dc}=16\rm \, T$ and zero field for respective configurations.}
\label{SlikaOP}
\end{figure}

Fig. \ref{SlikaOP} shows the curves measured in slightly underdoped YBCO single crystal in zero magnetic field, as well as in various applied magnetic fields. In an applied magnetic field, the superconducting  transition is shifted to a lower temperature, and occurs through the formation of the mixed state with some magnetic flux penetrating into the sample. The microwave currents then exert a Lorentz force on vortices so that their oscillation constitutes an  additional mechanism of the microwave power  absorption.\cite{Gittleman, Coffey, Dulcic} 

The measurements shown in Figs. \ref{SlikaOP}(a) and (b) were carried out in the two modes as described in Section II. Note that the signal in the normal state is considerably larger when $H_{\rm mw}$ is parallel with the $ab$ plane (Fig. \ref{SlikaOP}(b)). The explanation can be found if one observes that, in this configuration, the induced current flows in a closed loop, partly in the $ab$ plane and partly along the $c$ axis. Since the ratio of the normal state values $\rho_c$ and $\rho_{ab}$ is much larger than the ratio of the sample dimensions (the respective current paths), the main contribution to the absorption signal comes from the current path along the $c$ axis. The detailed analysis of $\rho_{ab}$ and $\rho_c$ in all the three samples will be postponed to Section IV. 

Here, we focus on the superconducting fluctuations above $T_c$. It is evident in Figs. \ref{SlikaOP}(a) and (b) that, at higher temperatures, the curves measured in applied fields of 16 T, and 8 T merge with the curve which was measured when no external magnetic field was applied. This overlap holds even on an expanded scale when the signal noise can be seen directly. It is useful to quantify this statement by analyzing the data in Fig. \ref{SlikaOP}(a). The total signal $1/2Q$ above $T_c$ is in the range 40-50 ppm, whereas our noise level in that temperature range is about 0.2 ppm. Hence, we can positively detect a relative change in our signal which is greater than $(1/2)10^{-2}$. Noting that $1/2Q$ is related to the surface resistance $R_s$ by a constant geometric factor, and $R_s \propto \sqrt{\rho}$, one finds that our method can detect a relative change in the resistivity $\Delta \rho / \rho$ greater than $10^{-2}$. A typical value for slightly underdoped YBCO at 100 K is about 100 $\mu\Omega$cm,\cite{Ando:04} so that our method is sensitive to resistivity changes greater than 1 $\mu\Omega$cm. Within this limit, we observe no magnetoresistance in fields up to 16 T in the normal state above 100 K in nearly optimally doped YBCO. We note, however, that the transverse and longitudinal dc magnetoresistance of magnitude $\sim 10^{-6}$ has been observed up to $\sim 270$ K over a wide range of underdoped YBCO.\cite{Ando:02} The small longitudinal  magnetoresistance at high temperatures was ascribed to magnetic scattering which increases when the spin gap is suppressed by the Zeeman effect, while a much larger magnetoresistance closer to $T_c$ was ascribed to the superconducting fluctuations. Orbital magnetoconductivity in the normal state at higher temperatures was fitted very well with the $(aT^{2}+b)^{-2}$ dependence, while an upturn from this behavior was observed only when it reached the magnitude $\sim 10^{-4}$ at temperatures about 20 K above $T_c$.

We conclude that the difference of the zero field and 16 T curves in Fig. \ref{SlikaOP}(a) can be used to analyze the appearance of the superconducting fluctuations above $T_c$. It is shown in the inset of Fig. \ref{SlikaOP}(a) on an expanded scale. The noise signal of about 0.2 ppm is clearly seen. The deviation from zero can be estimated to occur below 95~K. Below this temperature the superconducting fluctuations appear, and give rise to an excess conductivity. In the ac case, the superconducting conductivity is complex, but the onset of fluctuations is dominated by the real part of this conductivity. Here we can state that the superconducting fluctuations extend from $T_c = 87~ \rm K$ to $T' = 95~\rm K$, i. e. in the range of 
only 8 K, as seen from the insets in Figs. \ref{SlikaOP} (a) and (b). It is important to note that  similar results were found in nearly optimally doped Hg1201 single crystals,\cite{Grbic} and 
La$_{2-x}$Sr$_x$CuO$_4$ ,\cite{Pozek} indicating that this feature might be universal for  nearly optimally doped HTSC.

\subsection{Deeply underdoped sample}

\begin{figure}[t]
\includegraphics[width=8cm]{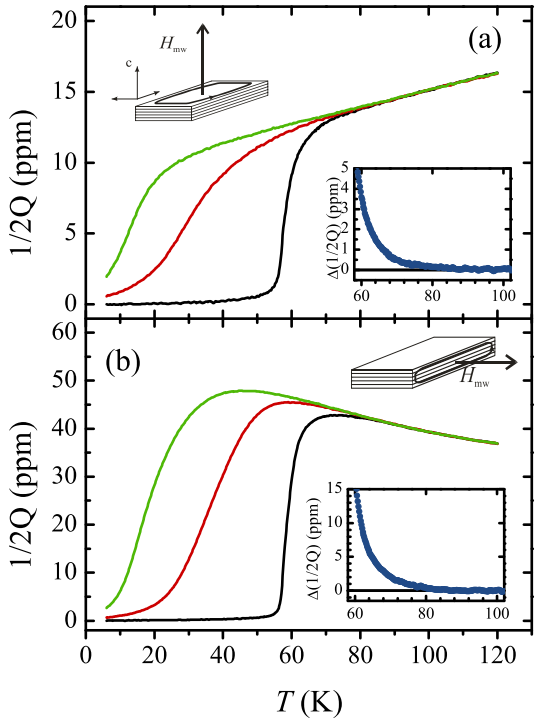}
\caption{(Color online) Microwave absorption in deeply underdoped YBCO ($T_c=57~\rm \,K$). External dc magnetic field is applied parallel to the $c$ axis (black: $B_{dc}=0$, red: $B_{dc}=8 \rm \, T$, green: $B_{dc}=16 \rm \, T$ ). (a) $H_{\rm mw}$ is in the $c$-direction so that the induced microwave currents flow in the $ab$ plane; (b) $H_{\rm mw}$ lies in the $ab$ plane so that the induced microwave currents flow partly in the $c$-direction and partly in the $ab$ plane. Insets show the difference in absorption between $B_{dc}=16\rm \, T$ and zero field for respective configurations.}
\label{SlikaUD}
\end{figure}

The microwave results in the deeply underdoped YBCO sample UD57 are shown in Fig. \ref{SlikaUD}. In the configuration where the microwave currents flow solely in the $ab$ plane (Fig. \ref{SlikaUD}(a)), one observes, even at such low doping level, metallic behavior in the normal state around 100 K, similarly as in the nearly optimally doped sample above. This is consistent with dc measurements of $\rho_{ab}$ in a series of YBCO samples.\cite{Ando:04} 

\begin{figure}[t]
\includegraphics[width=8cm]{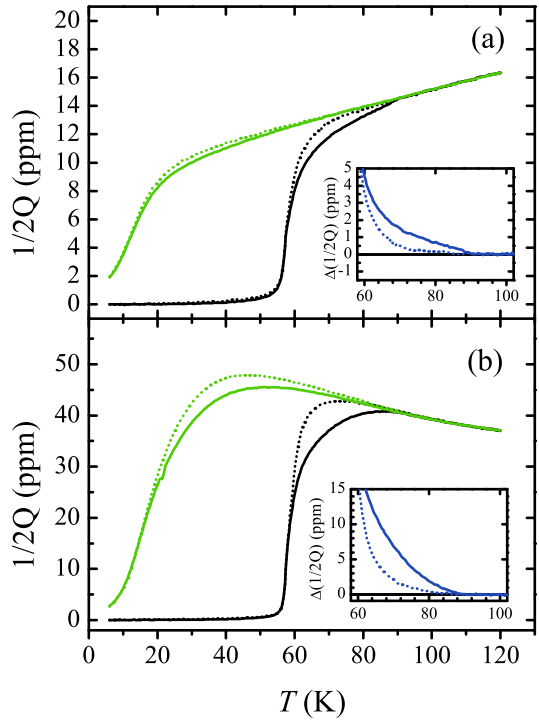}
\caption{(Color online) Microwave absorption in the deeply underdoped UD57 YBCO sample before (dotted lines), and after aging at room temperature in air (solid lines). The latter data show that the intake of oxygen takes place during aging, and a small fraction of the sample becomes optimally doped. External  dc magnetic field is applied parallel to the $c$ axis (black: $B_{dc}=0$, green: $B_{dc}=16 \rm \, T$ ). (a) $H_{\rm mw}$ is in the $c$-direction so that the induced microwave currents flow in the $ab$ plane; (b) $H_{\rm mw}$ lies in the $ab$ plane so that the induced microwave currents flow partly in the $c$-direction and partly in the $ab$ plane. Insets show the difference in absorption between $B_{dc}=16\rm \, T$ and zero field for respective configurations.}
\label{Aging}
\end{figure}

The transition to the superconducting state in Fig. \ref{SlikaUD}(a) is shifted to lower temperatures, and greatly broadened with respect to the slightly underdoped sample (compare Fig. \ref{SlikaOP}(a)).  The difference of the zero field and 16 T curves is shown in the inset of Fig. \ref{SlikaUD}(a) on an expanded scale. Some contribution of superconductivity is seen up to 80~K, i. e. 23~K above $T_c$. 
A similar temperature range of superconducting fluctuations in transport measurements was already reported for other HTSC materials, e. g. in underdoped samples of Bi$_{2}$Sr$_{2}$CaCu$_{2}$O$_{8}$  (Bi2212)\cite{Corson}, and Co-substituted YBCO.\cite{Bergeal} However, we should also consider the possibility that the sample might not be perfectly homogeneous, in which case some fractions of the sample would have $T_c$ values distributed slightly above 57~K, and reaching even 80~K. In this scenario, the superconducting fluctuations due to the main body of the sample might extend from 57~K to some temperature below 80~K, with their contribution to the superconducting conductivity being masked by that of a small-volume fraction of the sample having a distribution of $T_c$ values in that same temperature range. This, however, seems to be a minor effect in the freshly prepared sample. We demonstrate in Fig. \ref{Aging} that the microwave absorption can sensitively detect when a significant distribution of $T_c$ is introduced in the same sample after an extended period of aging (nearly five years in a closed box). Similar observations were reported earlier in  fluctuation diamagnetism of two YBCO samples (chain-ordered and chain-disordered) of the same $T_c$.\cite{Carretta} Our results for the extent of the superconducting fluctuations are consistent with those of Ref.~\onlinecite{Rullier:06} (Nernst effect) for a sample with the same $T_c$ before introducing disorder, and the result of Ref.~\onlinecite{LeBoeuf:07} (Hall effect). Within the sensitivity of our method, we can conclude that the observed superconducting fluctuations extend at maximum 23~K above $T_c$ in our UD57 YBCO sample.

The experimental curves presented in Fig. \ref{SlikaUD}(b) for the configuration in which $H_{\rm mw}$ lies in the $ab$ plane exhibit a different behavior than those in the nearly optimally doped sample (cf. Fig. \ref{SlikaOP}(b)). The semiconducting behavior in the normal state is obviously due to the contribution of the current flowing along the $c$ axis (inset in Fig. \ref{SlikaUD}(b)). The detailed behavior of $\rho_{ab}$ and $\rho_{c}$ is given later in Section IV.

Application of a high external magnetic field acts to suppress superconductivity close to $T_c$ so that the semiconductor-like behavior in Fig. \ref{SlikaUD}(b) is extended to lower temperatures. The difference in the signals measured in 16 T and zero field is shown in the inset to Fig. \ref{SlikaUD}(b) on an expanded scale. Again we find that the upper limit of the appearance of superconducting fluctuations is set at 80 K. 

\subsection{Overdoped sample}

\begin{figure}[t]
\includegraphics[width=8cm]{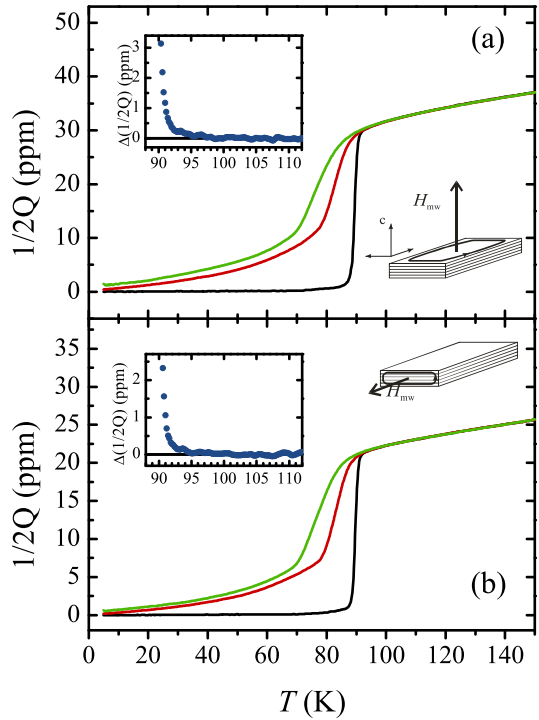}
\caption{(Color online) Microwave absorption in overdoped YBCO ($T_c=89~\rm \,K$). External dc magnetic field is applied parallel to the $c$ axis (black: $B_{dc}=0$, red: $B_{dc}=8 \rm \, T$, green: $B_{dc}=16 \rm \, T$ ). (a) $H_{\rm mw}$ is in the $c$-direction so that the induced microwave currents flow in the $ab$ plane; (b) $H_{\rm mw}$ is parallel to the $a$ axis so that the induced microwave currents flow partly along the $c$-axis and partly along the $b$ axis. Insets show the difference in absorption between $B_{dc}=16\rm \, T$ and zero field for respective configurations.}
\label{SlikaOD}
\end{figure}

We now turn to the overdoped side of the YBCO phase diagram. The measurements of the OD89 sample are shown in Fig. \ref{SlikaOD}. At this doping level the crystal structure is fully oxygenated, which enhances the interlayer coupling and results in a coherent normal-state transport along the $c$ axis. \cite{Homes, Hussey:N03} 
The resistivity $\rho_{c}$ is greatly reduced with respect to the underdoped samples, and the anisotropy $\rho_{c}/\rho_{ab}$ becomes small. As a result, the relative contribution of $\rho_{c}$ to the signal in Fig. \ref{SlikaOD}(b) is less pronounced than in the corresponding measurements of the underdoped samples. In Section V, we show that it was still possible to separately extract $\rho_{ab}$ and $\rho_{c}$ in the overdoped sample. Here we analyze the raw data curves by the same procedure as in the underdoped samples above. The difference of the zero field and 16 T curves shows that even in the overdoped sample the superconducting fluctuations are visible up to $T' = 96~\rm K$, i.e. 7~K above $T_c$.   

\subsection{Interpretation of the experimental data in the fluctuation regime}

Given the controversy that persists in the current literature about the temperature range of the superconducting fluctuations above $T_c$ in HTSC, we find it important to interprete our raw experimental data before performing any analysis which involves theoretical models, and/or assumptions on unknown behavior in the normal state. Our measurements of microwave absorption in progressively higher applied magnetic fields have shown that an 8 T field is already sufficient to supress superconductivity at temperatures above the zero-field $T_{c0}$ (cf. Fig. \ref{SigmeOP88field} in Section V). Here we use the curves measured in a 16 T field to represent the normal state at all temperatures above the zero-field $T_{c0}$, and observe the difference between these curves and those measured in zero applied field. Our results show that superconducting fluctuations extend 7-8~K above the zero-field $T_{c0}$ in overdoped and slightly underdoped YBCO samples, while in deeply underdoped YBCO sample the superconducting fluctuations extend at most up to 23~K above the zero-field $T_{c0}$. The latter result is the most intriguing in the current controversy about the nature of the pseudogap in deeply underdoped HTSC.

According to the theoretical analysis made by Emery and Kivelson,\cite{Emery, Emery2} the superconducting order parameter in underdoped HTSC samples has a low phase stiffness (superfluid density  of the system $\rho_s$), so that long range phase coherence can be achieved only below some temperature $T_\theta$, which is lower than the mean field transition temperature $T^{MF}$. The temperature $T_\theta$ then plays the role of the superconducting transition $T_c$ which, in a real system, may be additionally lowered by the effects of some other degrees of freedom. Thus, above $T_c$ this theory predicts the predominance of phase fluctuations extending in a wide temperature region. Some of the previous experimental observations may have observed this behavior.\cite{Corson, Grbic, Rullier:06} Furthermore, this theory predicts that even above the phase fluctuation region, and up to $T^{MF}$, one may have uncorrelated Cooper pairs, i. e. those which are not capable of forming even short range and short lived phase coherence.\cite{Lee:09} The crucial question is how high above $T_c$ can the superconducting fluctuations survive, and what experimental techniques can be used to positively detect them.

An external magnetic field perturbs the phase and amplitude of the superconducting order parameter. Which of the two components will be affected in the first line, depends on the phase stiffness of the system. The present experiments detect superconducting fluctuations which appear with some amplitude and phase correlation. In this sense, they are analogous to Josephson tunneling experiments reported previously.\cite{Bergeal} It is interesting to compare the frequency range of the detected fluctuations. Our experiments probe the superconducting fluctuations in the GHz frequency range, and we have shown that these fluctuations extend up to at most 23 K above $T_c$ for the UD57 sample. The excess conductance measured in the Josephson tunneling experiment as a function of the applied voltage revealed the superconducting fluctuations up to 1 THz frequency range at temperatures 14 K above $T_c$ in a Co-substituted YBCO sample with a similar doping level.\cite{Bergeal} These are conventional amplitude and phase superconducting fluctuations which are shown to persist within a relatively narrow temperature interval above $T_c$.

However, it is worth mentioning here that the measurements of the far-infrared (FIR) optical conductivity can probe the electron system at much higher frequencies.\cite{Yu} By considering the temperature dependence of the optical spectral weight, a recent FIR study in deeply underdoped YBCO (similar to our UD57 sample) has uncovered possible evidence for superconducting fluctuations up to temperatures as high as 180 K.\cite{Dubroka} The result was obtained using 
a multilayer model of the $c$-axis electrodynamics, which separates the local intra- and inter-bilayer conductivity in YBCO. Only the intra-bilayer conductivity was found to develop precursor conductivity at elevated temperatures. It gives rise to a transverse plasma mode, and anomalous temperature dependence of some infrared-active phonon modes. One may recall that neutron scattering experiments on underdoped YBCO have revealed spin correlations well above $T_c$ with fluctuation rates in the THz range.\cite{Hinkov:07, Hinkov:08} These spin correlations also develop a characteristic in-plane anisotropy, which has been interpreted in terms of an electronic transition into a nematic liquid state. This process may be related to the intra-bilayer anomalies observed in FIR measurements. 

On the other hand, microwave $c$-axis conductivity can detect SC fluctuations capable of forming long range inter-bilayer correlations. This can be further confirmed by comparing the samples with higher doping levels, where the correlations extend in the same temperature range when detected by both, microwave and FIR techniques. Namely, for the optimally doped and overdoped cuprates, model calculations \cite{Emery,Ussishkin} show that the phase stiffness is more rigid ($T_\theta > T^{MF}$), so that the transition to the normal state is expected to occur through the disappearance of the amplitude of the order parameter. Hence, above $T_c$ the superconducting fluctuations would preserve phase correlation as long as there is some amplitude present, 
i. e. amplitude fluctuations alone cannot be expected on theoretical grounds. The present measurements in  our slightly underdoped (UD87) and overdoped (OD89) samples have detected the superconducting fluctuations in the range of 7-8 K above $T_c$, which is a considerably shorter range than the one observed in the deeply underdoped (UD57) sample, but does not reduce to the narrow range characteristic of classical superconductors. These observations are  similar to the ones noted in recent measurements of the Nernst coefficient  in YBCO single crystals at different magnetic fields.\cite{Daou} Also, dynamic scaling analysis of the microwave conductivity in LSCO samples with a broad range of doping has shown a narrow range of fluctuations above $T_c$.\cite{Kitano:06} The same conclusion about the narrow range of fluctuations above $T_c$ has been recently drawn from THz time-domain spectroscopy in LSCO thin films with a doping range that spans almost the entire superconducting dome.\cite{Bilbro}

\section{Anisotropic resistivities}

Having established the extent of the superconducting fluctuations above $T_c$ directly from the measured data, we can proceed with a theoretical analysis of the measured curves, and reach additional conclusions which go beyond those of direct observation. 

In the course of microwave measurements, one determines two parameters, $1/2Q$ and $\Delta f/f$, where $\Delta f$ is the shift of the resonant frequency at a given temperature with respect to the lowest temperature value. The two parameters are combined to 

\begin{equation}
\frac{1}{2Q} - i \frac{\Delta f}{f} = \Gamma (R_s + i \Delta X_s)
\label{shift}
\end{equation}

\noindent
where $\Gamma$ is a geometric factor for a given sample in the cavity, which is not changed upon temperature variation, i. e. it holds for the entire set of data. It can be calculated if the resistivity $\rho_{n}$ is known at some temperature well above $T_c$ so that the sample is fully in the normal state. The calculated value of the normal surface resistance $R_{sn} = \sqrt{\omega \mu_{0} \rho_{n}/2}$ is then simply  compared to the measured value of $1/2Q$ at that temperature. 

One has to set an offset value to the measured frequency shift in Eq. (\ref{shift}) so that the full value of the surface reactance $X_s$ also appears in that equation. This can be accomplished through knowledge of the low temperature London penetration depth $\lambda_{L}$, from which one calculates $X_{s}(0) = \omega \mu_{0} \lambda_{L}$, or by assuming that the conductivity in the normal state is a real quantity $\sigma_{n}$ so that $X_{sn} = R_{sn} = \sqrt{\omega \mu_{0} \rho_{n}/2}$. We adopt the latter procedure for the analysis of our data in YBCO samples. The measured complex frequency shift can then be expressed through the surface impedance $\tilde{Z}_{s} = R_s + i X_s$ at all temperatures.

\begin{figure}[t]
\includegraphics[width=8cm]{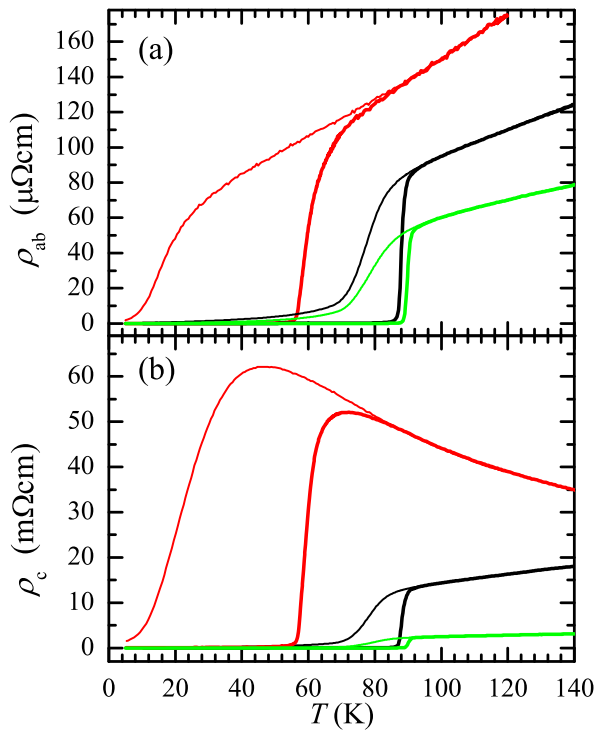}
\caption{(Color online) (a) The real part of the $ab$-plane resistivity of the three YBCO samples, OD89 (green), UD87 (black), and UD57 (red) as determined from the microwave measurements in the mode in which the current flows only in the $ab$-plane (Figs. \ref{SlikaOP}(a), \ref{SlikaUD}(a), and \ref{SlikaOD}(a)). Metallic behavior is observed in overdoped and underdoped samples in the normal state. The thicker and thiner lines pertain to the results obtained in zero field and 16 T fields, respectively. (b) The real part of the $c$-axis resistivity determined by the decomposition of the total signal measured in the mode where the current flows partly in the $ab$-plane and partly along the $c$-axis as shown in the insets of Figs. \ref{SlikaOP}(b), \ref{SlikaUD}(b), and \ref{SlikaOD}(b). The decomposition procedure is described in the text.}
\label{otpornosti_graf}
\end{figure}

The real part of the resistivities $\rho_{ab}$ for the three YBCO samples are shown in Fig. \ref{otpornosti_graf} (a). They were infered from the measurements shown in Figs. \ref{SlikaOP}(a), \ref{SlikaUD}(a), and \ref{SlikaOD}(a) where the sample orientation and the selected microwave mode ensured that the current was flowing only in the $ab$ plane. The in-plane resistivities are metallic-like, even in the deeply underdoped sample, as has been previously found by dc resistivity measurements.\cite{Ando:04} 

The resistivity $\rho_c$ can be extracted from the curves shown in Figs. \ref{SlikaOP}(b), \ref{SlikaUD}(b), and \ref{SlikaOD}(b) which are taken in the mode where the current flows partly in the $ab$ plane, and partly along the $c$ axis. For the microwave field along the $b$ axis, the effective measured surface impedance is 

\begin{equation}
Z_{eff} =  \frac{ab Z_{ab} + bc Z_{c}}{ab + bc} 
\label{impedance}
\end{equation}

\noindent
where $Z_{ab}$ and $Z_{c}$ are the true surface impedances depending on the resistivities $\rho_{ab}$ and $\rho_{c}$, respectively. The parameters $a$, $b$, and $c$ stand for the sample dimensions as given 
in Table \ref{tb1}. The resulting curves for the real part of $\rho_{c}$ are shown in Fig. \ref{otpornosti_graf} (b). The slopes of $\rho_{c}$ in overdoped and slightly underdoped samples exhibit a metallic-like behavior,  consistent with previously reported measurements of $\rho_{c}$ in optimally doped YBCO.\cite{Kitano:06, Hussey} In deeply underdoped UD57 sample, $\rho_{c}$ shows a semiconductor-like  behavior as previously reported by dc  resistivity,\cite{Takenaka} and microwave measurements\cite{Kitano:95} in underdoped YBCO. We note here that a similar semiconductor-like behavior of $\rho_{c}$ has been observed also in nearly optimally doped Hg1201 single crystals below the pseudogap temperature $T^*$.\cite{Grbic} However, the Hg1201 system is highly anisotropic already at optimal doping, while the YBCO system acquires a high anisotropy only when deeply underdoped. 

\section{Analysis of the complex conductivity due to superconducting fluctuations}

It is convenient to study the superconducting fluctuations through the complex conductivity $\tilde{\sigma} = \sigma_1 - i \sigma_2$, which can be infered from the complex surface impedance $\tilde{Z}_{s}$ using the expression  

\begin{equation}
\sigma_1 - i \sigma_2 = \frac{i \omega \mu_{0}}{Z_{s}^{2}}
\label{conductivity}
\end{equation}

\begin{figure}[t]
\includegraphics[width=8cm]{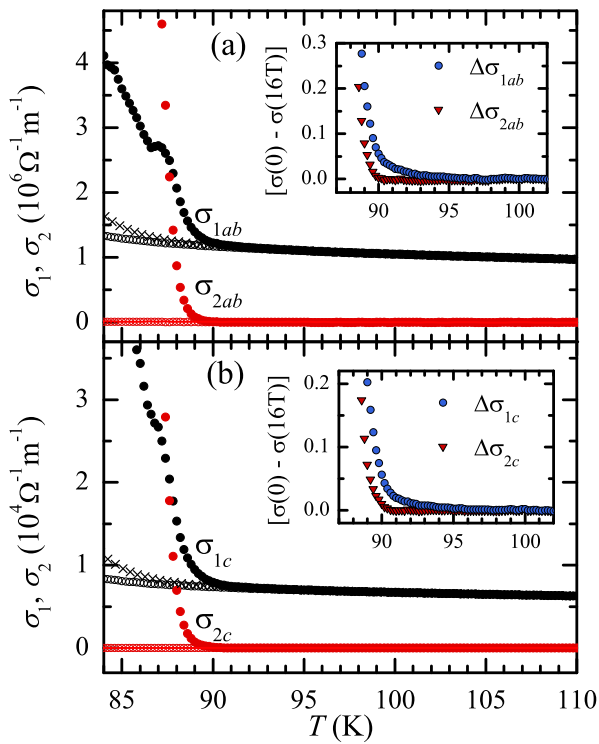}
\caption{(Color online) (a) $ab$-plane complex conductivity of the UD87 sample. The data points of the real part of the conductivity ($\sigma_{1ab}$) are represented by black symbols, while red symbols are used for the imaginary part ($\sigma_{2ab}$). Full circles, crosses, and open triangles stand for the conductivities in zero magnetic field, 8 T, and 16 T fields, respectively. The inset in (a) shows on an expanded scale the difference between the conductivities in zero field and in the field of 16 T ($\Delta\sigma_{1ab}$ blue circles, $\Delta\sigma_{2ab}$ red triangles) at temperatures above the zero-field $T_{c0}$. (b) The complex conductivity along the $c$ axis in the UD87 sample. The presentation is analogous to that of (a).}
\label{SigmeOP88field}
\end{figure}

The resulting real and imaginary parts of the zero field $ab$-plane conductivity in the slightly underdoped UD87 sample are shown in Fig. \ref{SigmeOP88field}(a). In ac measurements, the superconducting fluctuations bring about finite contribution to conductivities at  $T_c$.\cite{Schmidt, Dorsey} Moreover, the real part $\sigma_{1}$ has a peak, which may be superimposed on a broader maximum extending below $T_c$.\cite{Anlage, Waldram, Peligrad:04, Broun, Corson} The peak in $\sigma_{1}$ is a very useful feature since the zero field $T_c$ can be determined from a physically meaningful observation, rather than by referring to some arbitrary criteria, such as e. g. the midpoint on the measured transition curve in Fig. \ref{SlikaOP}(a). 

Also shown in Fig. \ref{SigmeOP88field}(a) are the $\sigma_{1ab}$ conductivities obtained from the curves measured in the applied magnetic fields of 8 T, and 16 T. It is evident that the latter two curves overlap at all temperatures above the zero field $T_c$. In other words, it is sufficient to apply the field of 8 T in order  to suppress all superconductivity above the zero field $T_c$, and the field of 16 T is seen to supress the superconductivity a few degrees further below the zero field $T_c$. Hence, by taking the difference of the conductivities in zero field and in 16 T field, one may construct a reliable procedure for extracting the pure superconducting fluctuation contribution to the conductivity above the zero field $T_c$. The inset in Fig. \ref{SigmeOP88field}(a) shows the result on an expanded scale. It is seen to merge into the signal noise at 95~K. This is the temperature $T'$ that has already been identified  directly form the microwave absorption data in Section III. 

The imaginary part of the fluctuation conductivity $\sigma_{2ab}$ is also presented 
in Fig. \ref{SigmeOP88field}(a). Note that the curves obtained from the measurements in 8 T, and in 16 T fields are indistinguishable in the whole temperature interval presented in Fig. \ref{SigmeOP88field}(a). They show zero values, as expected for the imaginary part of the conductivity in the normal state. Hence, the values of $\sigma_{2ab}$ obtained in zero field are due to the superconducting fluctuations. These data points are 
also shown in the inset of Fig. \ref{SigmeOP88field}(a) on an expanded scale. The imaginary part $\sigma_{2ab}$  is seen to decay much faster than the real part $\sigma_{1ab}$, which is expected at our operating microwave frequency.\cite{Peligrad:03} 

Fig. \ref{SigmeOP88field}(b) shows the complex conductivity along the $c$ axis in the UD87 sample. The presentation is analogous to that of the in-plane conductivity in panel (a). The signature of $T_c$ is visible as a shoulder on the curve of $\sigma_{1c}$. We note that both curves, $\sigma_{1ab}$ and $\sigma_{1c}$, yield the same value of $T_c$. It is the temperature at which the coherence lengths $\xi_{ab}$ and $\xi_{c}$ diverge. The applied field of 8 T is seen to suppress the superconductivity along the $c$-axis  at all temperatures above $T_c$, similarly as it has been shown in the $ab$ plane. Therefore, the curves measured in the field of 16 T can be safely used to represent the normal state at all temperatures above the zero field $T_c$. In the inset of Fig. \ref{SigmeOP88field}(b) one can observe that the fluctuation contribution to  $\sigma_{1c}$ extends up to 95~K, the temperature also identified as $T'$ in Section III. 

We have conducted a similar analysis in our deeply underdoped UD57, and overdoped OD89 samples. The general features are common to all the three samples. We find that the peak in the real part $\sigma_{1}$ of the complex conductivity, both in the $ab$ plane, and along the $c$ axis, can be used to determine the zero field $T_c$ in each of the samples. Also, the applied field of 16 T is found to be more than sufficient to suppress superconducting fluctuations above $T_{c0}$, so that the excess fluctuation conductivity above $T_c$ in zero field can be determined by the above described procedure. It appears that the temperature $T'$, up to which the real part $\sigma_{1}$ of the fluctuation conductivity extends before merging with the noise signal, coincides with the values already established in Section III. 

\begin{figure}[t]
\includegraphics[width=8cm]{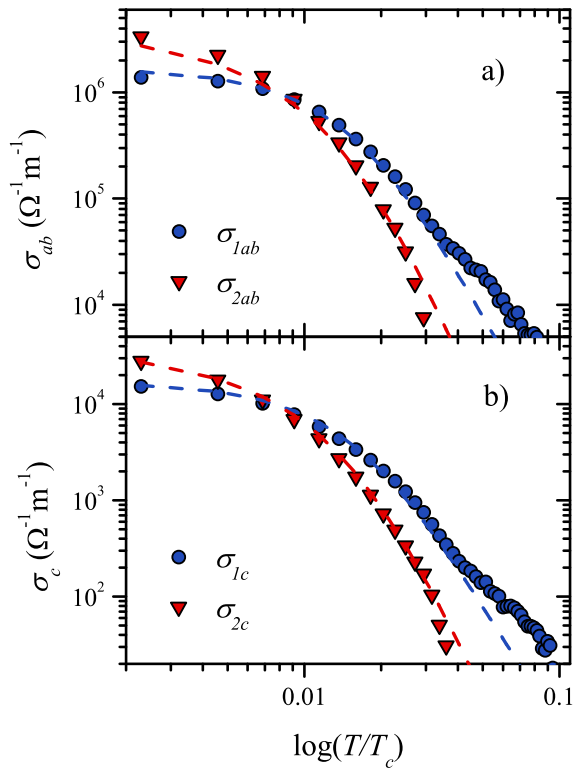}
\caption{(Color online) Presentation of the real and imaginary parts of the complex superconducting fluctuation conductivity above $T_c$ as a function of reduced temperature $t = \ln(T/T_c)$ on a log-log plot. Data for the slightly underdoped sample UD87 is shown by full symbols, and the dashed lines are theoretical curves as described in the text. (a) In-plane conductivity $\sigma_{1ab}$ and $\sigma_{2ab}$. (b) Conductivity along the c axis $\sigma_{1c}$ and $\sigma_{2c}$.}
\label{Fluct_sigmeOP87}
\end{figure}

It is useful to pay additional attention to the features of the real and imaginary component of the ac conductivity which arise due to the superconducting fluctuations. The data points for the in-plane superconducting fluctuation conductivity above $T_c$ in our slightly underdoped UD87 sample are presented in Fig. \ref{Fluct_sigmeOP87} against the reduced temperature $t = \ln(T/T_c)$ in a log-log plot. One may try to compare those experimental results to the theoretical predictions for the fluctuation conductivity in anisotropic three-dimensional (3D) systems. The test is severe since a single set of parameters is expected to describe well both the real and imaginary components of the fluctuation conductivity. 

The frequency dependent conductivity can be calculated within the Kubo formalism from the current due to the fluctuations of the order parameter,\cite{Tinkham, Schmidt} or, alternatively, as the response of the system to an external field. \cite{Dorsey, Wickham} It was also shown that, for physical reasons, one had to truncate the summation over the fluctuation modes at a minimum wavelength of $2\pi \xi_{0}/\Lambda$, where $\xi_{0}$ is the intrinsic coherence length, and $\Lambda$ is the cutoff parameter.\cite{Patton, Gollub} Indeed, with no cutoff taken in the calculations, the theoretical dc fluctuation conductivity greatly overestimates the experimentally observed values at temperatures above $T_c$.\cite{Hopfengartner} For the in-plane complex ac fluctuation conductivity in anisotropic 3D systems, one obtains\cite{Peligrad:03, Silva:02, Silva:04}

\begin{equation}
\tilde{\sigma}_{ab} = \frac{e^{2}}{32 \hbar \xi_{c0}}\left(\frac{\xi_{ab}(T)}{\xi_{ab0}}\right)^{z-1}\left[S_{1}(\omega, T, \Lambda) - 
i S_{2}(\omega, T, \Lambda)\right]
\label{fluctuation_cutoff}
\end{equation}

\noindent
where the intrinsic coherence length $\xi_{c0}$ along the $c$ axis enters in the prefactor. Here, the time dependent Ginzburg-Landau theory is used with the relaxational dynamics implying that the dynamical critical exponent $z = 2$. This value of $z$ was found previously by other experimental groups,\cite{Osborn, Kitano:06} and elaborated on theoretical grounds.\cite{Aji} However, alternative values for the dynamical critical exponent have also been quoted in the literature.\cite{Booth, Nakielski} Recently, it has been shown that the original experimental results in thin films can be analyzed properly only by accounting for the finite size effects, and the value of $z$ then extracted.\cite{Xu:09} In microwave measurements, the phase of the complex fluctuation conductivity at $T_c$ is often taken as a sensitive probe for the value of $z$. However, the extracted value of $z$ may still be incorrect if the effect of the short wavelength cutoff at $T_c$ is not accounted for properly.\cite{Peligrad:03} For those reasons, we retain the relaxational dynamics (model A) with $z = 2$, and determine the cutoff parameter from the fit to the experimental data. 

The temperature dependence is expressed through the reduced coherence length $\xi_{ab}(T)/\xi_{ab0}$ in both, the prefactor and in the $S_{1,2}$ functions as given explicitly elsewhere.\cite{Peligrad:03, Silva:02} Well above $T_c$ one expects the Gaussian regime with $(\xi_{ab}(t)/\xi_{ab0})=t^{-\nu}$, where $t=(T-T_c)/T_c$ is the reduced temperature, and the static critical exponent $\nu = 1/2$. However, in the critical region close to $T_c$, the Gaussian regime no longer holds, and one may include the quartic term in the Ginzburg-Landau functional using the Hartree approximation yielding finally a renormalized coherence length\cite{Dorsey, Neri}

\begin{equation}
\frac{\xi_{ab}(T)}{\xi_{ab0}} = 
\frac{\Gamma}{\ln(T/T_c)}\left(1 + \sqrt{1 + \frac{\ln(T/T_c)}{\Gamma^{2}}}\right)
\label{crossover}
\end{equation}

\noindent
where $\Gamma$ is a parameter which determines the crossover from the Gaussian temperature dependence ($\nu = 1/2$) to the critical regime ($\nu = 1$) close to $T_c$. Here, $T_c$ is the renormalized critical temperature, as observed in the actual experiment.

The fluctuation conductivity along the $c$ axis takes also the form of Eq. (\ref{fluctuation_cutoff}), except that $\xi_{c0}^{-1}$ in the prefactor is replaced by $\xi_{c0}/\xi_{ab0}^{2}=1/\gamma^{2}\xi_{c0}$, where $\gamma = \xi_{ab0}/\xi_{c0}$ is the anisotropy factor.\cite{Silva:02} The renormalized coherence length $\xi_{c}(T)/\xi_{c0}$ should be equal to $\xi_{ab}(T)/\xi_{ab0}$ given in Eq. (\ref{crossover}) if the anisotropy does not change with temperature. 

Fig. \ref{Fluct_sigmeOP87} shows the calculated ac fluctuation conductivity at our operating frequency of 15.15 GHz. We note first that $T_c$ is determined unambiguously from the peak in $\sigma_{1}$ shown in Fig. \ref{SigmeOP88field}, hence it is not a free parameter for the fit. This is a salient convenience in the microwave method, as compared to e. g. dc conductivity analysis where $T_c$ has to be considered as a fit parameter. Also, $\xi_{c0}$ in Eq. (\ref{fluctuation_cutoff}) can be determined by evaluating $\sigma_{2ab}$ at $T_c$, and equating the result to the experimental value.\cite{Peligrad:03, Peligrad:04} In the case of our UD87 sample, we have obtained  $\xi_{c0}=0.08~\rm nm$. The only free parameters are the cutoff parameter $\Lambda$ in Eq. (\ref{fluctuation_cutoff}), together with the crossover parameter $\Gamma$ in Eq. (\ref{crossover}). The best agreement of the theoretical curves and the experimental data for $\sigma_{ab}$ in Fig. \ref{Fluct_sigmeOP87}(a) is obtained with $\Lambda=0.032$, and $\Gamma=0.19$. The fit is also very good for $\sigma_{c}$ in Fig. \ref{Fluct_sigmeOP87}(b) with the same values of the crossover parameter $\Gamma$, and the cutoff parameter $\Lambda$. The anisotropy parameter was found to be $\gamma=10$.

\begin{figure}[t]
\includegraphics[width=8cm]{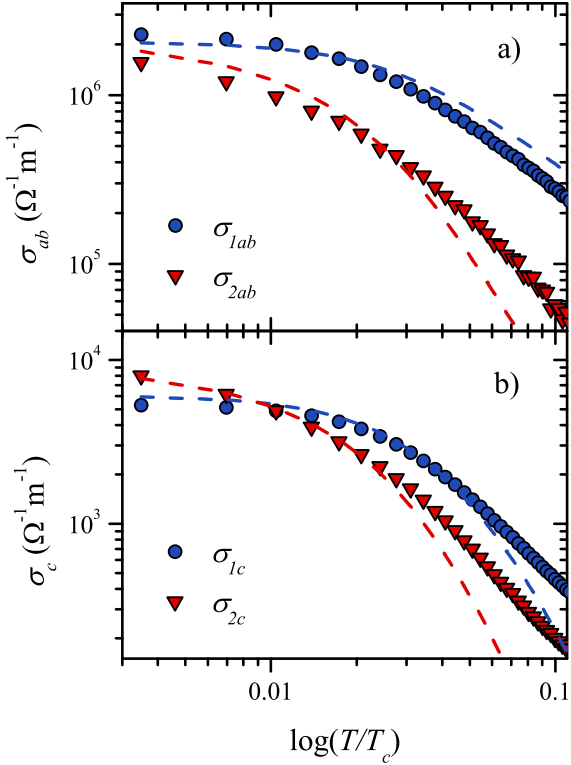}
\caption{(Color online) Real and imaginary parts of the complex superconducting fluctuation conductivity above $T_c$ in deeply underdoped sample UD57. Experimental values are shown by full symbols, and the dashed lines are theoretical curves as described in the text. (a) In-plane conductivity $\sigma_{1ab}$ and $\sigma_{2ab}$. (b) Conductivity along the c axis $\sigma_{1c}$ and $\sigma_{2c}$.}
\label{Fluct_sigmeUD57}
\end{figure}

A similar analysis of the fluctuation conductivity in our UD57 sample is presented in Fig. \ref{Fluct_sigmeUD57}. Again, $T_c$ is determined from the peak in $\sigma_{1}$, and is not a free parameter in the theoretical calculations. The crossover parameter is free, but we required it to be the same for in-plane and $c$-axis conductivities. The obtained value $\Gamma=0.4$ is larger than that found in the slightly underdoped sample UD87, which means that the critical fluctuations have a relatively larger extent in the deeply underdoped sample UD57. The cutoff parameters are different for the in-plane conductivity ($\Lambda=0.4$), and along the $c$ axis ($\Lambda=0.06$). This is expected in the underdoped sample with larger anisotropy. From the values of the conductivities at $T_c$ we found in the UD57 sample the coherence length $\xi_{c0}=0.13~\rm nm$, and the anisotropy factor $\gamma=15$. The overall agreement of the theoretical curves with the experimental data is good also in the deeply underdoped sample UD57, but it is not as remarkable as in the slightly underdoped sample UD87 shown in Fig. \ref{Fluct_sigmeOP87}. These tests are very sensitive to the homogeneity of the sample. It is known that even a small distribution of $T_c$ values brings about a noticeable reduction of the imaginary part $\sigma_{2}$.\cite{Kitano:06} The results shown in Fig. \ref{Fluct_sigmeUD57} indicate that even the freshly prepared sample UD57 contained a minor distribution of $T_c$ values, however at this point it is not obvious whether this inhomogeneity is intrinsic to a certain doping level as has been found in Bi$_{2}$Sr$_{2}$CaCu$_{2}$O$_{8+\delta}$,\cite{Gomes} and La$_{2-x}$Sr$_{x}$CuO$_{4+\delta}$.\cite{Iguchi} Regardless of its origin, this observation does not change our main conclusion in Section III that the superconducting fluctuations do not extend beyond 80~K, i.~e. they are confined within a relatively narrow region above $T_c$, and far below the pseudogap temperature $T^*$ in this system. We also note that magnetization measurements in optimally doped YBCO have shown good agreement with the theoretical expectations, but a deviation appeared in a deeply underdoped sample,\cite{Carretta} possibly also due to some distribution of $T_c$ values. 

The superconducting fluctuations appear simultaneously in both, the $ab$ plane and along the $c$ axis, already from the uppermost starting temperature. This observation questions some claims that, due to an increased anisotropy in underdoped samples, electronic coupling along the $c$ axis is completely lost in this system, so that only a KTB transition could take place.\cite{Ioffe, Herbut:04, Rullier:07} Our results are also at variance with the recent microwave observations in LSCO thin films, where 2D behavior was found in underdoped, and in overdoped samples, while 3D behavior was established only in the optimally doped sample.\cite{Ohashi}

\begin{figure}[t]
\includegraphics[width=8cm]{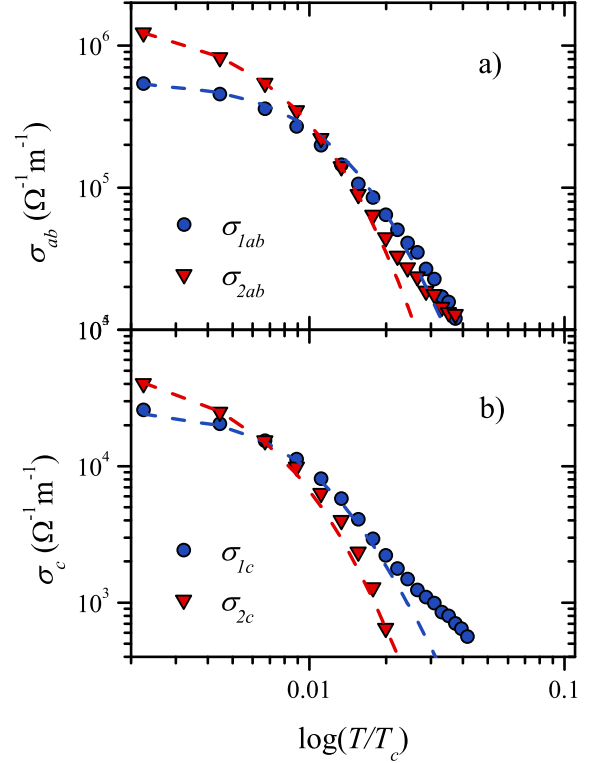}
\caption{(Color online) Real and imaginary parts of the complex superconducting fluctuation conductivity above $T_c$ in overdoped sample OD89. Experimental values are shown by full symbols, and the dashed lines are theoretical curves as described in the text. (a) In-plane conductivity $\sigma_{1ab}$ and $\sigma_{2ab}$. (b) Conductivity along the c axis $\sigma_{1c}$ and $\sigma_{2c}$.}
\label{Fluct_sigmeOD89}
\end{figure}

Fluctuation conductivities have been analyzed also in our overdoped sample, and the results are shown in 
Fig. \ref{Fluct_sigmeOD89}. The values of $\sigma_{2ab}$ and $\sigma_{2c}$ at $T_c$ yielded parameters $\xi_{c0}=0.17~\rm nm$ and anisotropy factor $\gamma=5.5$. The fit parameters were $\Gamma=0.21$ and $\Lambda=0.026$ for the pair of fluctuation conductivities $\sigma_{1ab}$ and $\sigma_{2ab}$. The fits to the pair $\sigma_{1c}$ and $\sigma_{2c}$ were obtained with similar values $\Gamma=0.16$ and $\Lambda=0.031$. The small discrepancy in the values of the crossover parameter $\Gamma$ could be due to the anisotropic conductivity which occurs in the $ab$ plane in overdoped samples, while our microwave measurement yielded an averaged value.

\section{Conclusions}

We have presented microwave absorption measurements of  YBCO single crystals in a broad doping range. By progressively increasing the applied magnetic field, we have shown that the field of 16 T is well sufficient to suppress all superconducting fluctuations above the zero field $T_c$. Hence, the difference of the curves measured in zero field, and in the field of 16 T is shown to be a sensitive and reliable method to detect the appearance of the superconducting fluctuations above $T_c$. This method does not rely on theoretical assumptions, and/or models of extracting the unknown parameters from the measured data. 

Our study shows that the temperature range of the superconducting fluctuations in YBCO is doping dependent. In the underdoped sample with $T_c=57$ K, it extends at most up to 23 K above $T_c$. If a minor distribution of $T_c$ values is present, the true extent of the superconducting fluctuations would appear even somewhat smaller. In any case, this relatively wide range of superconducting fluctuations is qualitatively consistent with the theoretical predictions for phase fluctuations in underdoped samples,\cite{Emery, Emery2, Lee, Franz, Herbut, Anderson} but provide strong evidence that the extent of the fluctuations falls well below the pseudogap temperature $T^*\approx 270$~K.  

The present results, however, do not exclude the possibility of intra-bilayer precursor superconducting fluctuations which might be observed by a high frequency probe, such as  FIR optical conductivity measurements, at intermediate temperatures within the pseudogap region of underdoped YBCO samples.\cite{Dubroka} 

In the slightly underdoped ($T_c=87~\rm K$), and overdoped ($T_c=89~\rm K$) samples, the phase stiffness is high enough to keep the phase correlation present whenever there is some local amplitude of the order parameter. Hence, we interpret the fluctuations observed in these samples by the present method as conventional fluctuations of the complex two-component superconducting order parameter. The extent of the fluctuations is found to be 8~K and 7~K in the nearly optimally doped and overdoped samples, respectively. 

We have also shown that the real and imaginary parts of the fluctuation conductivity determined at our microwave frequency are in compliance with theoretical calculations which take into account the proper cutoff for the wavevector pertaining to the superconducting fluctuation modes, and renormalized behavior of the coherence length in the critical region near $T_c$.

\begin{acknowledgments}
We would like to thank I. Kup\v ci\' c, E. Tuti\v s, S. Bari\v si\'c, A. Dubroka, C. Bernhard, and C. Berthier for numerous and invaluable discussions. This work was supported by The Croatian Ministry of Science through grant 119-1191458-1022, and by the German Science Foundation under grant FOR538.
\end{acknowledgments}

\end{document}